\documentstyle[epsfig,rotate,aps,preprint]{revtex}
\begin{document}
\draft
\author{A.T. Kruppa$^1$, K. Varga$^{1,2}$ and J. Revai$^3$}
\address{
$^1$Institute of Nuclear Research of the Hungarian Academy of Sciences, H-4001
Debrecen, Pf. 51, Hungary\\ $^2$Physics Division, Argonne National Laboratory,
Argonne, Illinois 60439, USA\\ $^3$ KFKI Research Institute for Particle and
Nuclear Physics, H-1525, Budapest, P.O.B. 49, Hungary}

\title{Local realizations of contact interactions in 
two- and three-body problems}
\maketitle

\begin{abstract}
Mathematically rigorous theory of the two-body contact interaction
in three dimension is reviewed.
Local potential realizations of this proper contact interaction are
given in terms of P\"oschl-Teller, exponential and square-well potentials.
Three body calculation is carried out for the
halo nucleus $^{11}$Li using adequately represented contact interaction.
\end{abstract}

\pacs{PACS numbers: 21.45.+v, 21.60-n, 21.30.-x, 27.20.+n}
\section{Introduction}

The contact interaction, or as sometimes called point or zero range
interaction, plays an important role in nuclear physics. The short range
nucleon-nucleon interaction is often simulated by a contact interaction
in different models. An effective interaction of contact type
(Skyrme-force) has been very frequently and successfully used in
mean-field descriptions of heavy nuclei
including the Hartree-Fock or Hartree-Fock-Bogoliubov theory
(see e.g. \cite{Do96}). Recently contact interaction has been applied
in three-body calculations of halo nuclei \cite{Be91,Es92,Es97} and
in the study of the role of resonance states in pair correlations \cite{Be96}.
Another example is the constituent-quark model 
for the application of the contact interaction.  
Several
non-relativistic three-body calculations have calculated the spectra
of the three quark system \cite{Gl96,Gl98,Fe98}.
Despite the popularity of the  contact force even its definition
is problematic. The Hamiltonian of a two-body system interacting
via a contact interaction is written as
\begin{equation}\label{2bdelta}
\hat H_\delta=
-{\hbar^2\over 2\mu}\Delta_{{\bf r}} +\lambda \delta ({\bf r}),
\end{equation}
where $\mu$ is the reduced mass, $\Delta_{{\bf r}}$
is the three dimensional Laplacian, $\delta ({\bf r})$
is the Dirac delta-function and $\lambda$ is the ``strength" of the
contact interaction. The interpretation of the
equation (\ref{2bdelta}) is not trivial due to the singular nature
of the Dirac delta-function. By using a complete set of states
to diagonalize the above Hamiltonian one gets infinite binding energy
for the system. To get a more meaningful result one usually truncates
the configuration space \cite{Be91,Es92,Es97,Be96,Fe98} with some
``reguralization'' scheme. This truncation leads to a model dependence 
and results of different approaches can only be related and reproduced
if one uses the same strategy to reguralize the interaction. Due to this
approach it is widely believed that the contact
interaction is meaningful only in a truncated space.

In this paper we propose a different approach where the application
of the contact interaction is based on mathematically rigorous theory
\cite{Al88}. This proper contact interaction (PCI) is free from the
above mentioned problems. The PCI gives finite energy and there is
no dependence on basis truncations or extra parameters. This PCI has
zero range and it acts non-trivially only in the $l=0$ partial wave.
These two properties are what physicists expect from a contact interaction.

The contact force was introduced in nuclear physics by Bethe and Peierls
\cite{Be35} and Thomas \cite{Th35} in 1935.
The mathematically correct treatment of the operator (\ref{2bdelta}) dates
back to the work of Berezin and Faddeev \cite{Be61} in 1961.
An English review of the early Russian works on two and three-body problems
with PCI can be found in Ref. \cite{Fl67}.
Mathematically rigorous treatment of the operator (\ref{2bdelta})
is given by different approaches.
One of them is the coupling constant
re-normalization procedure of Berezin and Faddeev \cite{Be61}. This technique
was applied also in two dimensions \cite{Ma94}. Other approaches are
the reguralization of the Green-function \cite{Gr94,Pa95},
the dimensional reguralization \cite{Mi98}, or the
theory of the Dirichlet forms \cite{Al80}.
We will use the self-adjoint extension theory of the contact force
which is described in great detail in the monograph \cite{Al88}.
Each of these approaches leads to the same spectrum of the two-body
Hamiltonian with PCI.

The mathematically correct formulation of the PCI in two-body problem
is completely solved. It gives the two-body t-matrix in analytical
form.
This t-matrix can be used in subsequent three-body calculations based
on the Faddeev equations. However, if we want to solve the three-body
Schr\"odinger equation in coordinate space using some approximations then we need a local
realization of the PCI. The aim of the paper is to give a few explicit
local potential realizations of the PCI and apply it
in three-body calculations.

The plan of the paper is the following. In section II the two-body
spectrum of the PCI is discussed. In part III the general theory of
the local potential realization of the PCI is reviewed and a few
explicit realizations are given.
In sections II and III.A those theorems of the monograph \cite{Al88}
are reviewed which are relevant for our purpose. In section IV the
numerical solution of the two-body problem with PCI are shown.
Finally, at
the section V we present three-body calculation with PCI for the halo
nuclei $^{11}$Li.

\section{Contact interaction in two-body problem}

Our review of the mathematically correct
treatment of the contact interaction
is based on the monograph of Albeverio {\it
et al.} \cite {Al88}.
The self-adjoint extension
approach can be originated from the intuitive picture of the
Dirac delta-function.
The ``function" $\delta({\bf r})$ is zero if ${\bf r}$
is different from the zero vector ${\bf 0}$. This means that
the ``operator" $\hat H_\delta$ is ``identical"
with the kinetic energy $-{\hbar^2\over 2\mu}\Delta_{{\bf r}}$
if ${\bf r}\ne {\bf 0}$. In mathematical terms
this means that the operator $-{\hbar^2\over 2\mu}\Delta_{{\bf r}}$
acts only  on the functions which belong to the function
space $C_0^\infty({\cal R}^3\setminus{\bf 0})$.
This restricted Laplacian is not self-adjoint but its symmetric closure
has many self-adjoint extensions. The self-adjoint
extensions are characterized by a real number $\alpha$
($-\infty<\alpha<\infty$).
This self-adjoint extension, $\hat H_{\alpha,\delta}$, will give 
the proper representation of the contact interaction.

The spectrum of the operator $\hat H_{\alpha,\delta}$ is the 
following. The continuous spectrum is $[0,\infty)$ for each $\alpha$.
If $\alpha< 0$ then the $l=0$ partial wave component of 
$\hat H_{\alpha,\delta}$ has one bound state with energy
\begin{equation}\label{benergy}
E_{\alpha,\delta}=-\left(4\pi\alpha\right)^2{\hbar^2\over 2\mu}.
\end{equation}
The Hamiltonian $\hat H_{\alpha,\delta}$ has no bound states for 
higher partial waves.
If the parameter of the self-adjoint extension is non-negative then
there is no bound state in any partial wave but
there is one virtual state in the partial wave $l=0$ with
energy given by Eq. (\ref{benergy}).

The continuous spectrum may be characterized by the phase shift
$\delta_l^{\alpha,\delta}(k)$, where $k$ denotes the wave
number $k=(2\mu E/\hbar^2)^{1/2}$ and $E>0$. In the partial wave $l=0$ the
phase shift reads
\begin{equation}\label{effr}
k {\rm cot}(\delta_0^{\alpha,\delta}(k))=-{1\over a_{\alpha,\delta}}, 
\end{equation}
where
\begin{equation}\label{scatl}
a_{\alpha,\delta}=-{1\over 4\pi\alpha}.
\end{equation}
The higher partial wave phase shifts are identically zero.
The S-matrix in the partial wave $l=0$ has the following form
\begin{equation}\label{PCIS}
S^{\alpha,\delta}_0(k)={4\pi\alpha+ik\over 4\pi\alpha-ik}.
\end{equation}
The full characterization of a Hamiltonian is given if we know its
Green-function. The Green-function of PCI in coordinate space can be
given in closed analytical form \cite{Al88}.

The effective range expansion of an ordinary local spherical potential
is of the form $k^{2l+1}{\rm cot}\delta_l(k)=-1/a_l+r_lk^2/2+...$, where
$a_l$ is the scattering length and $r_l$ is the effective range.
The comparison of Eq. (\ref{effr}) with this
expansion shows that the operator $\hat H_{\alpha,\delta}$ contains
an interaction with zero range. The self-adjoint extension parameter
not only determines the binding energy (see Eq. (\ref{benergy})
if $\alpha<0)$) but it is also connected to the scattering length by Eq.
(\ref{scatl}).

At first it may seem strange that continuously many PCI exist.
One can understand this
using the following arguments (see e.g chapter 14.1 in
Ref. \cite{Ne82}).
If a spherical potential vanishes for $r\geq r_0$ then
the solution of the radial Scr\"odinger equation is determined
by assigning a given value to the logarithmic derivative of the
wave function at $r_0$. This boundary condition depends on the
energy. If the range of the potential $r_0$ tends to zero (contact force)
then the boundary condition parameter becomes independent of the energy
and in this way
we can assign any value to it. Thus continuously many
zero range interactions can be defined. This approach is an another
way to define the PCI.

To sum up, the
two-body problem with Dirac delta-potential can be solved
analytically in a mathematically rigorous way. Infinite binding energy
due to the contact force does not occur.
The mathematically correct treatment of the contact force
always leads to a finite and well defined results.

\section{Local potential realization of the contact force}

The abstract PCI can be realized by an appropriate limit of
local potentials. We have to construct this local representation in
such a way that it corresponds to the self-adjoint extension theory.
Naively taking a potential which resembles to a Dirac $\delta$ function will
not work.

\subsection {Realization of the contact interaction by local potential}
The recipe for local realization of the contact interaction
is given in the monograph \cite{Al88} (see theorem 1.2.5).
Let us take any local
potential $V({\bf r})$ which is in the Rollnik class
\cite{Al88,Ne82} and assume that
$\int d{\bf r}(1+r)\left\vert V({\bf r})\right\vert$ is finite. We
define a Hamiltonian family by
\begin{equation}\label{hamfam}
\hat H_\delta(\epsilon)=-{\hbar^2\over 2\mu}\Delta_{{\bf r}}
+{\lambda(\epsilon)\over \epsilon^2}V({\bf r}/\epsilon),
\end{equation}
where $\epsilon>0$ and $\lambda (\epsilon)$ is
an arbitrary analytic function of $\epsilon$ such that $\lambda(0)=1$.
Let us assume further that the potential $V({\bf r})$
has a non-degenerate zero energy resonance, i.e.
\begin{equation}\label{resonance}
-{\hbar^2\over 2\mu}\Delta_{{\bf r}} \Phi +V({\bf r})\Phi=0,
\end{equation}
for some $\Phi$. Because $\Phi$ is a resonance state
it is not square integrable function. The normalization is
defined by the following condition
\begin{equation}\label{norm}
\int d{\bf r} V({\bf r}) \left\vert\Phi ({\bf r})\right\vert^2=-1.
\end{equation}
Under these circumstances  the Green-operator of the
Hamiltonian family (\ref{hamfam})
converges to the Green-operator of $\hat H_{\alpha,\delta}$
in norm resolvent sense if $\epsilon\rightarrow 0$. The self-adjoint
extension parameter is given by
\begin{equation}\label{alpha}
\alpha =-\lambda'(0){\hbar^2\over 2\mu}\left (
\int d{\bf r} V({\bf r}) \Phi ({\bf r})\right )^{-2}.
\end{equation}
The theorem 1.2.5 of Ref. \cite{Al88} includes other
cases. Here we have quoted only the part that we will use.

\subsection{Realization by P\"oschl-Teller potential}

We will realize the result of the previous section
using the P\"oschl-Teller potential. It is known that
in this case the radial Scr\"odinger equation
\begin{equation}\label{rPT}
-{\hbar^2\over 2\mu}u''(r)-{\hbar^2\over 2\mu b^2}
{s(s+1)\over {\rm cosh}(r/b)^2}u(r)=E u(r),
\end{equation}
can be solved analytically \cite{Ba96}.
The energy eigenvalues are given by
\begin{equation}
E_n=-(s-1-2n)^2{\hbar^2\over 2\mu b^2},
\end{equation}
where $n=0,2,4\ldots$, and $s$ is an arbitrary complex
number. If we choose the strength so that $s=2\nu+1$,
where $\nu=0,2,4,\ldots$, then we will have solutions with zero energy
for the state $n=\nu$.
The corresponding wave functions are
\begin{equation}
u_\nu (r)\propto C_{2\nu+1}^{1/2}\left({\rm tanh}(r/b)\right),
\end{equation}
where $C^{1/2}_{2\nu+1}$ is the Gegenbauer polynomial.
It is easy to see that these solutions are not square integrable so we have
potentials which have zero energy resonance states. For simplicity
we will consider the case $\nu=0$. The $\Phi$ of Eq. (\ref{resonance})
with the normalization (\ref{norm}) now reads
\begin{equation}\label{PhiPT}
\Phi ({\bf r})=\left({2\mu b\over \hbar^2}\right)^{1/2}
{\sqrt 3\ {\rm tanh}(r/b)\over r\sqrt {8\pi}}
.
\end{equation}
The self-adjoint extension parameter can be calculated using
Eqs. (\ref{alpha})
and (\ref{PhiPT}), 
\begin{equation}\label{PTalpha}
\alpha=-{\lambda'(0)\over 6b\pi}.
\end{equation}

Let us introduce the Hamiltonian family corresponding to 
Eq. (\ref{hamfam}),
\begin{equation}\label{PT}
\hat H_{PT}(\epsilon)=
-{\hbar^2\over 2\mu}\Delta_{{\bf r}}-{\lambda(\epsilon)\over
b^2\epsilon^2}{\hbar^2\over 2\mu}{2\over {\rm cosh}^2(r/(b\epsilon))}.
\end{equation}
We will call this operator
the scaled P\"oschl-Teller realization of the PCI.
According to the previous
section the resolvent of (\ref{PT}) converges to the resolvent of
$\hat H_{\alpha,\delta}$ as
$\epsilon\rightarrow 0$. We thus have a
local realization of the PCI.

In this example the
convergence of the Hamiltonian family (\ref{PT}) in the limit
$\epsilon\rightarrow 0$
can be easily shown using elementary calculations.
Since the eigenvalue problem of the
scaled P\"oschl-Teller operator (\ref{PT}) can be written in
the form (\ref{rPT}) the eigenvalues of (\ref{PT}) read
\begin{equation}\label{PTe}
E_{PT}(\epsilon,n)=-\left({s(\epsilon)-1-2n\over b\epsilon}\right)^2
{\hbar^2\over 2\mu},
\end{equation}
where
\begin{equation}
s(\epsilon)={1\over 2}\left[-1+\left(1+8\lambda(\epsilon)\right)^{1/2}\right]
\end{equation}
and $n=0,2,4,\ldots$. Let us 
calculate $E_{PT}(\epsilon,n=0)$ in the limit $\epsilon \rightarrow 0$.
We get
\begin{equation}
\lim_{\epsilon\rightarrow 0}
E_{PT}(\epsilon,0)=\lim_{\epsilon\rightarrow 0}
-\left({s(\epsilon)-1\over \epsilon}\right)^2{\hbar^2\over 2\mu b^2}=
-s'(0)^2{\hbar^2\over 2\mu b^2},
\end{equation}
since in the limit $\epsilon\rightarrow  0$ we
have $s(\epsilon)\rightarrow 1$.
Calculating the derivative of $s(\epsilon)$ at $\epsilon=0$ we end up with
\begin{equation}
\lim_{\epsilon\rightarrow 0} E_{PT}(\epsilon,0)=-{4\over  9}\lambda'(0)^2
{\hbar^2\over 2\mu b^2}.
\end{equation}
Thus we get the same energy as in the abstract theory of the PCI, 
i.e., if we substitute
(\ref{PTalpha}) into (\ref{benergy}).

The $s$-wave S-matrix of the scaled P\"oschl-Teller potential is
of the following form \cite{Ba96}
\begin{equation}\label{delPT}
S_0^{PT}(k,\epsilon)=-\left({1\over 2}\right)^{2ikb\epsilon}
{\Gamma (ikb\epsilon)\Gamma({1\over 2}(2+s(\epsilon)-ikb\epsilon))
\Gamma({1\over 2}(1-s(\epsilon)-ikb\epsilon))
\over
\Gamma (-ikb\epsilon)\Gamma({1\over 2}(2+s(\epsilon)+ikb\epsilon))
\Gamma({1\over 2}(1-s(\epsilon)+ikb\epsilon))}.
\end{equation}
Using the identity $\Gamma(1+z)=z\Gamma(z)$ of the Gamma function
we can easily calculate the limit $\epsilon\rightarrow 0$ of the S-matrix
(\ref{delPT})
\begin{equation}
\lim_{\epsilon\rightarrow 0}S_0^{PT}(k,\epsilon)=
\lim_{\epsilon\rightarrow 0}{1-s(\epsilon)+ikb\epsilon\over
1-s(\epsilon)-ikb\epsilon}={s'(0)+ikb\over-s'(0)-ikb}.
\end{equation}
Taking into account Eq. (\ref{PTalpha}) and $s'(0)=2 \lambda'(0)/3$
we see that the limit agrees with the expression (\ref{PCIS}) given by the
general abstract theory of the PCI.

\subsection{Realization by exponential and square well potentials}

A local realization of the abstract result can
be given not only with
the P\"oschl-Teller potential but by any local potential satisfying the
conditions of section III.A. Here we take first the exponential potential
as an another example.
The eigenenergies of the radial Scr\"odinger equation,
\begin{equation}\label{rEXP}
-{\hbar^2\over 2\mu}u''(r)-V_0\exp(-r/b)u(r)=E u(r),
\end{equation}
can be obtained from the transcendental equation \cite{Ne82}
\begin{equation}
J_{2ibk}\left(2b\sqrt {{2\mu V_0\over \hbar^2}}\right)=0,
\end{equation}
where $J_\nu$ is the Bessel function.
The condition of the zero energy resonance reads
\begin{equation}\label{exp1}
J_0\left(2b\sqrt {{2\mu V_0\over \hbar^2}}\right)=0.
\end{equation}
This equation defines the strength of the potential. The corresponding
resonance wave function with the normalization given in section III.A is
\begin{equation}\label{phiexp}
\Phi({\bf r})={1\over r}{1\over \sqrt{4\pi bV_0}
J_1(b_0)}
J_0(b_0\exp(-r/2b)),
\end{equation}
where $b_0=2b(2\mu V_0/\hbar^2)^{1/2}$.

The scaled Hamiltonian family which
converges to the PCI in the limit
$\epsilon\rightarrow 0$ is
\begin{equation}\label{exp2}
-{\hbar^2\over 2\mu}\Delta_{{\bf r}} -V_0{\lambda(\epsilon)\over
\epsilon^2}\exp(-r/b\epsilon).
\end{equation}
Note that the potential strength $V_0$ is determined by
Eq. (\ref{exp1}).
We call the operator (\ref{exp2}) the scaled exponential
realization of the PCI.
The self-adjoint extension parameter can be calculated using
Eqs. (\ref{alpha}) and (\ref{phiexp})
\begin{equation}\label{exp3}
\alpha=-\lambda'(0) {\hbar^2\over 2\mu}
{bJ_1(b_0)^2\over 4\pi V_0}
\left[\int_0^\infty dr r \exp(-r/b)
J_0\left(b_0\exp(-r/2b)\right)\right]^{-2}.
\end{equation}
The integral in (\ref{exp3}) can be calculated numerically.

The abstract PCI can also be realized by a properly chosen square well
potential. This latter is defined in the usual way:
$V^{SQ}(r)=-V_0\ {\rm if}\ r<R$ and $V^{SQ}(r)=0\ {\rm if}\ r>R$.
The condition of the presence of a resonance with zero energy is that
the strength of the square well is
\begin{equation}
V_0={\hbar^2\over 2\mu}\left[{(2k+1)\pi\over R}\right]^2,
\end{equation}
where $k=0,\pm1,\pm2,\ldots$. The properly normalized zero energy
resonance wave function reads
\begin{equation}
\Phi({\bf r})={1\over r}(2\pi V_0 R)^{-1/2}
\sin \left(\sqrt{{2\mu\over \hbar^2}V_0}r\right).
\end{equation}
The self-adjoint extension parameter is given by
\begin{equation}
\alpha=-\lambda'(0) {2\mu\over \hbar^2}{RV_0\over 8\pi}.
\end{equation}
The Hamiltonian family which tends to the PCI is the following
\begin{equation}
-{\hbar^2\over 2\mu}\Delta_{{\bf r}}+
{\lambda (\epsilon)\over\epsilon^2}\hat V^{SQ}_\epsilon(r),
\end{equation}
where $V^{SQ}_\epsilon(r)=-V_0\ {\rm if}\ r<R\epsilon$ and
$V^{SQ}_\epsilon (r)=0\ {\rm if}\ r>R\epsilon$.

\section{Numerical calculation: two-body problem}

We have seen that the eigenvalue problem
of the P\"oschl-Teller realization of
the PCI can be solved
exactly for any $\epsilon$. Since we will
solve the three-body problem numerically, using the stochastic variational
method \cite{Su98}, it is essential to see how this numerical method performs
for the solution of Eq. (\ref{PT}) or any other local realization of
the abstract PCI.
This numerical exercise is indeed a new test for the
stochastic variational method since the potential of (\ref{PT})
really resembles the Dirac delta-function. 
A few scaled P\"oschl-Teller potentials are shown in Fig. 1.
for $b=1\ fm$.

For the function $\lambda(\epsilon)$ we have chosen the
following ansatz $\lambda(\epsilon)=1+C\epsilon$. With the parameter $C$
we can change $\lambda'(0)$
which influences the value of the
self-adjoint extension parameter.
This latter uniquely determines the bound state energy and
the scattering length of the PCI.

For the two body problem we considered the bound neutron-proton system
and the scattering states of the neutron-neutron system. For the
deuteron case
the exact eigenvalues and the numerically calculated ones
using the scaled P\"oschl-Teller potential are
compared in Table I for several values of
the scaling parameter $\epsilon$. We have used
$\hbar^2/(2\mu)=41.47\ {\rm MeV\ fm}^2$ in all calculations.
Table I demonstrates that the stochastic variational
method is able to deal with such an extreme potential.
Of course, numerically, we could not carry out the
limit $\epsilon\rightarrow 0$.
Using the reliable results of the stochastic variational method
$(\epsilon\leq0.005)$
the numerically calculated values can be extrapolated
to $\epsilon=0$. From Table I
we can conclude that the calculation with
$\epsilon=0.005$ is very satisfactory; the error is less than
$10$ $eV$.
Having this experience, 
in three-body calculations we will accept the result
of the calculation with scaling $\epsilon=0.005$ (or the
extrapolated value to $\epsilon=0$) as the result of the PCI.

The numerical results of the exponential and square well realization
of the PCI are given in Table II. 
The parameters of the potentials were $V_0=$0.936831 MeV, $b=$8 fm and
$V_0=$4.092925 MeV and $R=$ 5 fm in the case of the exponential and 
square well potentials, respectively.
The stochastic variational
calculation performs very well for the exponential potential.
It was not possible get better accuracy than four digits in the case
of a square well potential.  The convergence
of the exponential and square well realization
to the PCI is not as quick as in the case of the P\"oschl-Teller
realization.

The phase shift of the scaled P\"oschl-Teller potential
$\delta^{PT}_0(k,\epsilon)$
can be deduced from Eq. (\ref{delPT}). The quantity
$k \cot\delta_0^{PT}(k,\epsilon)$ is
displayed in Fig. 2 as the function of the wave number
for several values of the scaling parameter.
We see that decreasing the scaling parameter the result
converges to the phase shift of the PCI.
The self-adjoint extension parameter of the PCI in this
calculation is fixed to give
the experimentally observed scattering length ($-18.5$ fm) of the
dineutron. In order to get the effective range parameters of the
scaled P\"oschl-Teller potential we make the expansion:
\begin{equation}
k\cot \delta_0^{PT}(k,\epsilon)=ik(S^{PT}_0(k,\epsilon)+1)/
(S^{PT}_0(k,\epsilon)-1)\approx-1/a_{nn}+r_0^2k^2/2.
\end{equation}
At each scaling the scattering length and the
effective range obtained in this way 
are shown in Tables I and III for the scaled P\"oschl-Teller 
potentials. 

To sum up, in numerical calculations we can only decrease the effective
range to $0.01$ fm but this is already enough to get the binding 
energy with
an accuracy less than $10$ $keV$.

\section{Contact interaction in three-body problem}

We have seen that the contact interaction is very tricky even for the
two-body problem. Further complications can be expected when three
particles interact. If each pair interaction is a contact force then
the three-body problem can be solved exactly \cite{Be61,Fl67}.
The final result is, however, physically unsatisfactory due to the 
Thomas effect:
there is an infinite number of bound states and
the spectrum is not bounded from below \cite{Th35,Ad88}.
We mention that in this three-body problem
the self-adjoint extension type approach
have to be used to have a  meaningful three-body Hamiltonian.

In the present paper we consider a
three-body system where only one pair of the particles interacts through
a contact force and the other two pair interactions are given by short
range local potentials. The physical motivation of such a three-body
system is to model nuclei with halos, like $^6$He or $^{11}$Li.
The situation in this case is simpler.
If we want to include the PCI in this type of  three-body problem 
we can proceed in two different ways.

For the solution of the three-body problem we may use the Faddeev
theory. In this case one needs the Green-operator
of each subsystem of the three particles.
Since one pair of particles interact via PCI the
Green-operator
of this two body subsystem is known exactly and analytically. This
property leads to a simplification of the Faddeev equations.
It can be shown that this type of three-body problem is well defined
and physically meaningful.
The theoretical framework of this Faddeev type approach was
developed in \cite{Ba66}.

If we want to solve the three-body problem using the Scr\"odinger
equation in coordinate space then we are ready to use the PCI since
we have several local realizations of the PCI.
Since we are interested in bound state properties of three-body
systems we can use the Rayleigh-Ritz variational principle to
calculate the binding energies.
For the three body problem we have chosen a
simple model of the light halo nucleus
$^{11}$Li which presently is in the focus of many theoretical and
experimental investigations.

Let us consider a simple model: an inert core and two neutrons.
The three-body Hamiltonian using Jacobi coordinates has the form
\begin{equation}
-{\hbar^2\over 2\mu}\Delta_{{\bf r}}
-{\hbar^2\over 2M}\Delta_{{\bf R}}
+V_{n,n}({\bf r})+V_{n,c}({\bf r},{\bf R}).
\end{equation}
Here ${\bf r}={\bf r_1}-{\bf r_2}$ is the relative coordinate of the
two neutrons and ${\bf R}={1\over 2}({\bf r_1}+{\bf r_2})-{\bf r_3}$
is the vector which points from the 
third particle (core) to the center of mass of the dineutron.
As in Refs. \cite{Es92,Es97} we assume that the neutron-neutron
interaction is a zero range contact force. However, we will not
restrict the model space, as it was done in Refs. \cite{Es92,Es97},
in order to get finite energy but we will use the PCI.
For the neutron-neutron interaction we will take the
scaled P\"oschl-Teller and exponential realizations of the PCI.
A shallow neutron core interaction is taken  in order to avoid the problem
of the Pauli projection. The explicit
form of this potential is given in Ref. \cite{Es97}.

In Table IV two calculations with the P\"oschl-Teller realization of the
PCI are shown. In the first case
the parameterization
of the PCI corresponds to the experimental value
($-18.5$ fm) of the
scattering length of the free neutron-neutron system.
The second calculation corresponds to the case
when the parameter of the PCI is adjusted until the three-body
energy reproduces the experimental value
($S_{2n}=-0.295\pm0.035\ MeV$) of the
two-neutron separation energy of $^{11}$Li \cite{Yo93}. This
parameter of the PCI
leads to a $-7.32$ fm scattering length for the dineutron.
The mean square radius of the neutron-neutron and the core-dineutron
system is also calculated in this case.

From Table IV we can conclude that
the effective neutron-neutron interaction  is not consistent
with the free neutron-neutron scattering. We had to reduce
the scattering length of the dineutron system
in order to reproduce the experimental value of the two
neutron separation energy. Similar phenomena were observed in
the calculations of Refs. \cite{Be91,Es92,Es97} where the traditional
notion of the contact force was applied. The results of Table IV are 
very similar to the results of Ref. \cite{Es97} where the standard
method (restricted model space) was used to handle the contact interaction.
Although the results are qualitatively similar
there is a deep conceptual difference between the two calculations.
In our calculations with PCI we indeed used
a zero range force whereas in the works \cite{Es92,Es97},
due to the restriction of the model space, the effective range is much
larger than ours. In principle the PCI has zero range, in practice we
can reach 0.01 fm. In the calculations of Ref. \cite{Es92,Es97} the
effective range is roughly 2 fm and this explain the quantitative
difference of the two calculations.

We have repeated the three-body calculation, replacing
the P\"oschl-Teller realization by the
exponential realization of the PCI. The results are shown in
Table V. Of course, due to the different realization of
the PCI, the energy corresponding to a given $\epsilon$
are different in Table V and in the second column of Table IV.
What is important is that the extrapolated values,  which correspond to
the PCI, are very close to each other in both realizations.

Because we use the local realization of the PCI we have to carry out
the limit $\epsilon\rightarrow 0$ numerically
in the three-body calculation and this requires great care.
The stochastic variational method \cite{Su98}, which proved to be very
accurate in few body problems, is used
to solve the bound state three-body problem.
It is clear from Tables IV and V that convergence is reached in
the order of a few 
tenth keV. This means that our proposed method for the correct handling
of the contact interaction works for three-body problems.

\section{Summary}

We have reviewed the mathematically, rigorously defined
proper contact interaction 
and shown that it has the properties of the standard notion
of the contact interaction. This proper contact interaction is free
from the problem (infinite binding energy) of the standard
treatment of the
contact interaction. Local potential realizations of the abstract proper
contact interaction were given in terms of P\"oschl-Teller,
exponential and square-well potentials.
The calculation for $^{11}$Li revealed that the proper
contact interaction with local realization
can be incorporated into three-body calculations based on the 
coordinate space Schr\"odinger equation 
provided this latter is solved accurately
enough.

\acknowledgments

The work of K. V.  was supported by the U. S. Department of Energy, 
Nuclear Physics Division, under contract No. W-31-109-ENG-39 and 
this research was also supported by Hungarian 
OTKA grants No. T029003 and T026244.

\newpage

\begin{table}
\caption{The binding energy in MeV of the deuteron in scaled P\"oschl-Teller
potential as the function of the scaling parameter. The exact result and the
numerical one calculated by the stochastic variational method
are shown. The binding energy of the proper
contact interaction is $2.222496$ MeV this
corresponds to the limit $\epsilon\rightarrow 0$. The effective range
$r_0$ of the scaled P\"oschl-Teller potentials are also given.}
\begin{tabular} {cccc}
$\epsilon$&$r_0\ (fm)$&$-E_{PT}$ numerical&$-E_{PT}$ exact\\
\hline
0.500&0.86&2.065933&2.065933\\
0.100&0.19&2.188844&2.188843\\
0.050&0.10&2.205511&2.205510\\
0.010&0.02&2.219079&2.219073\\
0.005&0.01&2.220789&2.220783\\
\hline
extrapolated to&\\
$\epsilon=0$&0.00&2.222502&2.222496
\end{tabular}

\end{table}

\begin{table}
\caption{The binding energy in $\rm MeV$ of the deuteron in scaled
exponential and square-well
potentials as the function of the scaling parameter. The exact result and the
numerical one calculated by the stochastic variational method
are shown. The binding energy of the proper
contact interaction is $2.222496$ MeV this
corresponds to the limit $\epsilon\rightarrow 0$.}
\begin{tabular} {ccccc}
&\multicolumn{2}{c}{exponential}&\multicolumn{2}{c}{square-well}\\
\hline
$\epsilon$&numerical&exact&numerical&exact\\
\hline
0.500&1.34894&1.34894&1.662&1.663\\
0.100&1.93918&1.93919&2.077&2.080\\
0.050&2.06796&2.06797&2.135&2.149\\
0.010&2.18910&2.18909&2.196&2.207\\
0.005&2.20562&2.20562&2.201&2.215\\
\hline
extrapolated to&\\
$\epsilon=0$&2.22250&2.22250&2.205&2.222
\end{tabular}

\end{table}

\begin{table}
\caption{The effective range parameters of the dineutron system
in scaled P\"oschl-Teller
potential as the function of the scaling parameter.
The effective range parameters of the proper
contact interaction (scattering length $a_{nn}=-18.5$ fm and
effective range $r_0=0$ fm)
corresponds to the limit $\epsilon\rightarrow 0$.}
\begin{tabular} {cccccc}
$\epsilon$&0.500&0.1&0.05&0.01&0.005\\
\hline
$-a_{nn}\ (fm)$&19.85&18.38&18.43&18.49&18.49\\
$r_0\ (fm)$&1.04&0.20&0.10&0.02&0.01
\end{tabular}

\end{table}

\begin{table}
\caption{Two neutron separation energy of $^{11}$Li in the three-body model
as the function of the scaling parameter of the P\"oschl-Teller potential.
The binding energy of the proper
contact interaction can be obtained extrapolating to $\epsilon=0$.
The self adjoint extension parameter of the proper contact
interaction corresponds to the given neutron-neutron scattering length
$a_{nn}$. The mean square neutron-neutron and core-dineutron
radius are also given.}
\begin{tabular} {c|c|ccc}
&$a_{nn}=-18.5\ {\rm fm}$&\multicolumn{3}{c}{$a_{nn}=-7.32\ fm$}\\
\hline
$\epsilon$&$S_{2n}\ (MeV)$&$S_{2n}\ (MeV)$&
$r_{nn}^2$ (fm$^2$)&$r_{cn}^2$ (fm$^2$)\\
\hline
0.500&-0.5495&-0.1590&86.3&113.0\\
0.100&-0.6949&-0.2812&58.0&79.6\\
0.050&-0.7135&-0.2975&55.1&76.0\\
0.010&-0.7268&-0.3095&54.3&75.6\\
0.005&-0.7285&-0.3112&54.1&75.3\\
\hline
extrapolated to&\\
$\epsilon=0$&-0.7302&-0.3130&53.9&74.9
\end{tabular}

\end{table}
\begin{table}
\caption{Two neutron separation energy of $^{11}$Li in the three-body model
as the function of the scaling parameter of the exponential potential.
The binding energy of the proper
contact interaction can be obtained extrapolating to $\epsilon=0$.
The self adjoint extension parameter of the proper contact
interaction corresponds to the free neutron-neutron scattering length
$-18.5\ {\rm fm}$.}
\begin{tabular} {cc}
$\epsilon$&$S_{2n}\ (MeV)$\\
\hline
0.100&-0.2786\\
0.050&-0.4797\\
0.010&-0.6795\\
0.005&-0.7058\\
0.0025&-0.7188\\
\hline
extrapolated to&\\
$\epsilon=0$&-0.7316
\end{tabular}

\end{table}

\newpage

\begin{figure}
\epsfxsize=7.7cm \begin{rotate}[r]{\epsfbox{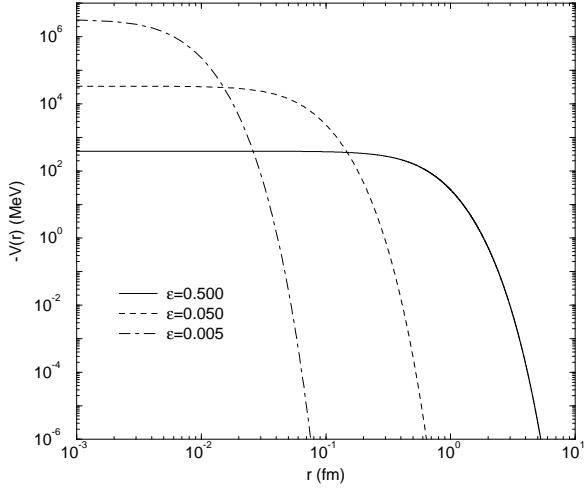}}\end{rotate}
\vskip 0.6cm
\caption{Local potential realization of the proper
contact interaction in the case of deuteron.
Scaled P\"oscl-Teller potentials
with different scaling parameters. Note the logarithmic scale
on each axis.}
\end{figure}
\begin{figure}
\epsfxsize=7.7cm \begin{rotate}[r]{\epsfbox{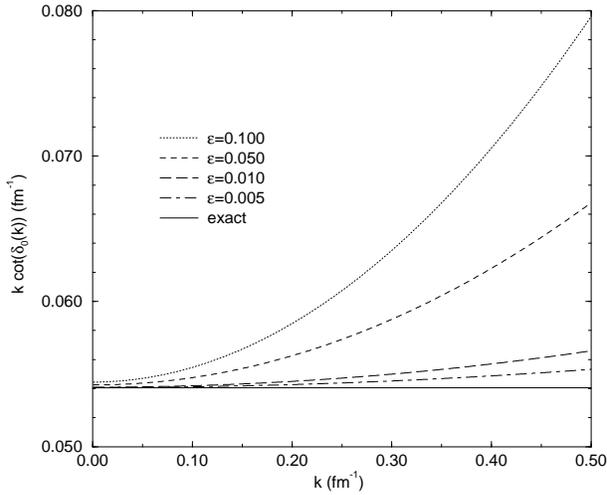}}\end{rotate}
\vskip 0.6cm
\caption{The phase shift of the scaled P\"oscl-Teller potential
with different scaling parameters for the dineutron system. The
parameter of the PCI corresponds to the experimental
scattering length $a_{nn}=-18.5\ fm$. The quantity $k\cot(\delta_0(k))$
is displayed. For the proper contact interaction this is independent
of the wave number $k$ (
$k\cot(\delta_0(k))={1\over 18.5}\ fm^{-1}$) }
\end{figure}

\end{document}